\documentclass[]{aa}
\usepackage{graphics}
\usepackage{graphicx}
\usepackage{txfonts}
\usepackage{natbib}
\bibpunct{(}{)}{;}{a}{}{,}

\def\aj{AJ}                   
             
\def\apj{ApJ}                 
                
\def\apjs{ApJS}               
           
\def\apss{Ap\&SS}             
\def\aap{A\&A}

\def\mnras{MNRAS}

\def\zap{ZAp}                 
\def\nat{Nature}

\def\nphysa{Nucl.~Phys.~A}

\def\msol{M$_{\odot}$}

\begin{document}      
%   \thesaurus{12.03.1;12.03.3;12.03.4;12.04.2;12.12.1;11.03.1}  
% 
   \title{Lithium abundances in exoplanet-host stars : modelling} 
% 
%   \subtitle{} 
 
        \titlerunning{Lithium abundances in exoplanet-host star}  
 
   \author{
M. Castro \ 
\inst{1}, 
S. Vauclair, 
\inst{2}
O. Richard
\inst{3}
\and N. C. Santos
\inst{4}
}
 
   \offprints{M. Castro}

   \institute{Centro de Astronomia e Astrof\'isica da Universidade de Lisboa, Observat\'orio Astron\'omico de Lisboa, Tapada da Ajuda, 1349-018 Lisboa, Portugal
\and Laboratoire d'Astrophysique de Toulouse et Tarbes - UMR 5572 - Universit\'e Paul Sabatier Toulouse III - CNRS, 14, av. E. Belin, 31400 Toulouse, France
\and Universit\'e Montpellier II - GRAAL, CNRS - UMR 5024, place Eug\'ene Bataillon, 34095 Montpellier, France
\and Centro de Astrof\'isica da Universidade do Porto, Rua das Estrelas, 4150-762 Porto, Portugal} 

\mail{mcastro@oal.ul.pt}
   
\date{Received \rule{2.0cm}{0.01cm} ; accepted \rule{2.0cm}{0.01cm} }

\authorrunning{M.Castro et al.}

\abstract
{}
{Exoplanet-host stars (EHS) are known to present superficial chemical abundances different from those of stars without any detected planet (NEHS). EHS are, on the average, overmetallic compared to the Sun. The observations also show that, for cool stars, lithium is more depleted in EHS than in NEHS. The aim of this paper is to obtain constraints on possible models able to explain this difference, in the framework of overmetallic models compared to models with solar abundances.} 
{We have computed main sequence stellar models with various masses and metallicities. The results show different behaviour for the lithium destruction according to those parameters. We compare these results to the spectroscopic observations of lithium.}
{Our models show that the observed lithium differences between EHS and NEHS are not directly due to the overmetallicity of the EHS: some extra mixing is needed below the convective zones. We discuss possible explanations for the needed extra mixing, in particular an increase of the mixing efficiency associated with the development of shear instabilities below the convective zone, triggered by angular momentum transfer due to the planetary migration.}
{}  
 

\maketitle

\section{Introduction}
				   
More than one year after the launching of the satellite CoRoT and at the time we receive the first data, the understanding of the formation of the planetary systems becomes crucial. In the present state-of-the-art, the differences in the protostellar clouds and in the internal structures between exoplanet-host stars (EHS) and stars without any detected planet (NEHS) are not well-known. Many studies \citep{gonzalez98,santos00,santos01,santos03} have shown that EHS are on average more metallic than NEHS. Two scenarii concerning the formation of planetary systems have been proposed to account for this behavior: an original overmetallic protostellar cloud or a process of accretion of overmetallic matter (planetesimals, asteroids, comets, etc.) onto the star. This second scenario is now ruled out for several reasons listed below.

Precise spectroscopic ground-based observations of lithium \citep{israelian03,ryan00,israelian04,chen&zhao06} and beryllium \citep{santos02,santos04b} in EHS and NEHS provide new constraints for stellar modelling. The study of the light element abundances may give us additional information about the rotation and angular momentum history of stars, particularly for the EHS. A different angular momentum history, and consequently different light-element abundances can be expected for EHS and NEHS, due to the presence of a more massive protostellar disc and possible accretion of planetary-mass bodies onto the star \citep{siess&livio99}. 

\citet{israelian04} show that a significant difference in the lithium abundance between EHS and NEHS in the range of effective temperature 5600-5850 K does exist. In this range of effective temperatures, EHS are more lithium underabundant than NEHS.

In this paper, we compute various stellar models to compare the theoretical lithium abundances to those determined from the spectroscopic observations. In the first part, we present the metallic peculiarities of EHS. In the second part, the evolution of various types of models calculated with the Toulouse-Geneva Evolution Code (TGEC) are presented and compared to the observed stars of \citet{israelian04} sample. \citet{bouvier08} has discussed the possibility of a different rotation history in EHS and NEHS to explain the lithium differences. In our models, rotation is treated in the same way for all the stars. In the third part, we discuss the influence of the presence of planets on the mixing in EHS, as suggested by \citet{israelian04} and \citet{chen&zhao06} and we present other models. Conclusions are given in the fourth section.

\section{Chemical peculiarities of exoplanet-host stars}

\subsection{Overmetallicity of exoplanet-host stars}

The most important difference between NEHS and EHS is the overmetallicity of the later \citep{gonzalez98,santos00,santos01,santos03}. The mean overmetallicity of EHS is $\sim$0.2 dex and the metallicity difference with NEHS is positive in 80\% of the cases. 

Two scenarii have been proposed to explain this metal enrichment in central stars of planetary systems. The first scenario assumes a protostellar gas initially metal-rich \citep{pinsonneault01,santos01,santos03}. Following the ``traditional'' view, the gaseous giant planets are formed by accretion of gas on planetesimals of around 10 earth masses. The higher the metallicity (and the more dust particles there are), the more rapidly planetesimals are formed, and the higher is the probability to form giant planets before the dissipation of the protostellar disc. In this scenario, the star is overmetallic from the centre to the surface. \\

The other suggestion is that the overmetallicity results from the accretion of planets, planetesimals, asteroids, comets and dusts on the surface of the star \citep{murray01}. The mixing of matter occurring in the upper layers of the star homogenises the abundances, modifying the ratio between heavy elements, brought by the accreted objects, and hydrogen. If this mixing process was restricted to the convective zone, the metal enrichment would strongly depend on the stellar mass, which is not the case \citep{pinsonneault01,santos01}. However, \citet{vauclair04} showed that, due to thermohaline mixing, the element dilution is more important in hotter stars than in cooler ones and could possibly account for a similar overabundance in stars of different masses. However, the possibility that the entire overmetallic matter sink inside the star is not negligible, in which case no overmetallicity would be left at all in the stellar outer layers. In any case, the large amount of accreted matter necessary to obtain the observed overmetallicity in cool stars, typically a hundred earth masses of iron, seems too large to be realistic.

Another point is that the accretion scenario could only work if accretion still occurs while the star arrives at the beginning of the main sequence: the material accreted during the pre main-sequence phase is too much diluted in deep convective zones to lead to an observable effect. While the migration time scales are quite small, of order $10^5$ - $10^6$ years, the formation of giant planets can take a much longer time, as large as a few $10^7$ years \citep[e.g.][]{ida&lin08}: planets may still form and migrate while the star is at the end of the PMS phase, but most of the accretion phase is expected to occur before that. \\

Several studies have tried to differentiate the two scenarii by different manners, including asteroseismology \citep{bazot&vauclair04}. A deep analysis of the star mu Arae, observed with HARPS \citep{bouchy05} has been given by \citet{bazot05}. In spite of the excellent precision obtained for this star, it was not yet possible to disentangle the two possibilities. \citet{vauclair08} gave evidence from asteroseismology that the exoplanet-host star $\iota$ Hor, visible in the south hemisphere, has very probably been formed together with the Hyades cluster, inside the same overmetallic cloud. This gives further evidence that the accretion hypothesis as an explanation of the observed metallicity is ruled out.

\subsection{Lithium abundances in exoplanet-host stars}

In \citet{israelian04}, the authors present a comparison of the lithium abundances in EHS and NEHS. Two samples of stars are compared : a sample with 79 EHS and a comparison sample of 157 NEHS from \citet{chen01}. 

The lithium distributions in the samples show an important statistical excess of EHS having a lithium abundance 1.0 $< \log \epsilon (\rm Li) <$ 1.6, among the coolest stars. 

Lithium abundances of EHS with effective temperature between 5850 and 6350 K are similar to those of the NEHS sample of \citeauthor{chen01}, whereas at lower effective temperatures the EHS show a clear lithium underabundance relative to the comparison sample. The excess of lithium-poor EHS is concentrated in the range 5600 K $< \rm T_{eff} <$ 5850 K. This study was confirmed by \citet{chen&zhao06}. 

A different angular momentum history, due to the presence of a more massive protostellar disc in the case of EHS \citep{edwards93,wolff04} or to accretion of planetary-bodies into the star \citep{siess&livio99,israelian03}, can affect the lithium abundances in EHS. A relative overmetallicity of the EHS, either initial or resulting from stellar pollution, could also be a clue to the abundance differences.

\section{Modelling}\label{sec:modelling} 

The following study consists of comparing the evolution of lithium (a) in stellar models with a solar metallicity \citep[``old'' abundances as given by][]{grevesse&noels93}, (b) in overmetallic models (with an initial overabundance of metals from the centre to the surface). The model with a solar metallicity is used as a reference model (standard model).

\subsection{Models and calibration}

The evolution of models with various masses are computed with the Toulouse-Geneva Evolution Code \citep{richard96,huibonhoa07} and parameterized as follows : \\

\noindent \textit{Inputs physics}

\noindent The OPAL2001 \citep{rogers&nayfonov02} equation of state is implemented. We use the OPAL96 opacities tables \citep{iglesias&rogers96} completed by the \citet{alexander&fergussen94} low temperature opacities. For nuclear reactions rates, we use the analytical formulae of the NACRE \citep{angulo99} compilation, taking into account the three \textit{pp} chains and the CNO tricycle, with the \citet{bahcall&pinsonneault92} screening routine. Convection is treated according to the \citet{bohmvitense58} formalism of the mixing-length theory with $\alpha_p = l/H_p = 1.75$. For the atmosphere, we use a grey atmosphere following the Eddington relation.\\

\noindent The abundance variations of the following chemical species are individually computed in the stellar evolution code: H, He, C, N, O, Ne and Mg. The heavier elements are gathered in Z. The initial composition follows the \citet{grevesse&noels93} mixture.\\

\noindent \textit{Diffusion and rotation-induced mixing}

\noindent All models include gravitational settling with diffusion coefficients computed as in \citet{paquette86}. Radiative accelerations are not computed here, as we only focus on solar-type stars where their effects are negligible.\\

\noindent Rotation-induced mixing is computed as described in \citet{theado&vauclair03}. This prescription is an extension of \citet{zahn92} and \citet{maeder&zahn98}, and introduces the feedback effect of the $\mu$-currents in the meridional circulation, due to the diffusion-induced molecular weight gradients. The evolution of the rotation profile follows the \citet{skumanich72} law. The models have a surface initial rotation velocity on the ZAMS equal to $V_i = 100$ km.s$^{-1}$. Other prescriptions, which include the angular momentum transport induced by the mixing, have been given by \citet{charbonnel&talon99} and \citet{palacios03}. However, as rotation-induced mixing alone cannot account for the flat rotation profile inside the Sun, these authors have later introduced the possible effect of internal gravity waves triggered at the bottom of the convective zone \citep[e.g.][]{talon&charbonnel05}. Other authors suggest that the internal magnetic field is more important than internal waves to transport angular momentum \citep{gough&mcintyre98}. In any case, when adjusted on the solar case, all these prescriptions are able to reproduce the lithium depletion observed in the Hyades, and the results are finally quite similar \citep{talon&charbonnel98,theado&vauclair03}. \\

\noindent We also include a shear layer below the convective zone, treated as a tachocline \citep[see][]{spiegel&zahn92}: this layer is parameterized with an effective diffusion coefficient decreasing exponentially downwards \citep[see][]{brun98,brun99}:
\begin{displaymath}
D_{tacho} = D_{bcz} \exp \left( \ln 2 \frac{r - r_{bcz}}{\Delta} \right)
\end{displaymath}
where $D_{bcz}$ and $r_{bcz}$ are respectively the value of $D_{tacho}$ at the bottom of the convective zone and the radius at this location, and $\Delta$ is a parameterized width. 

\noindent No overshooting is added here.\\

\subsection{Results}

\subsubsection{Influence of the metallicity on the lithium destruction}

\begin{figure}[h!]
\begin{center}
\includegraphics[angle=0,height=8cm,width=\columnwidth]{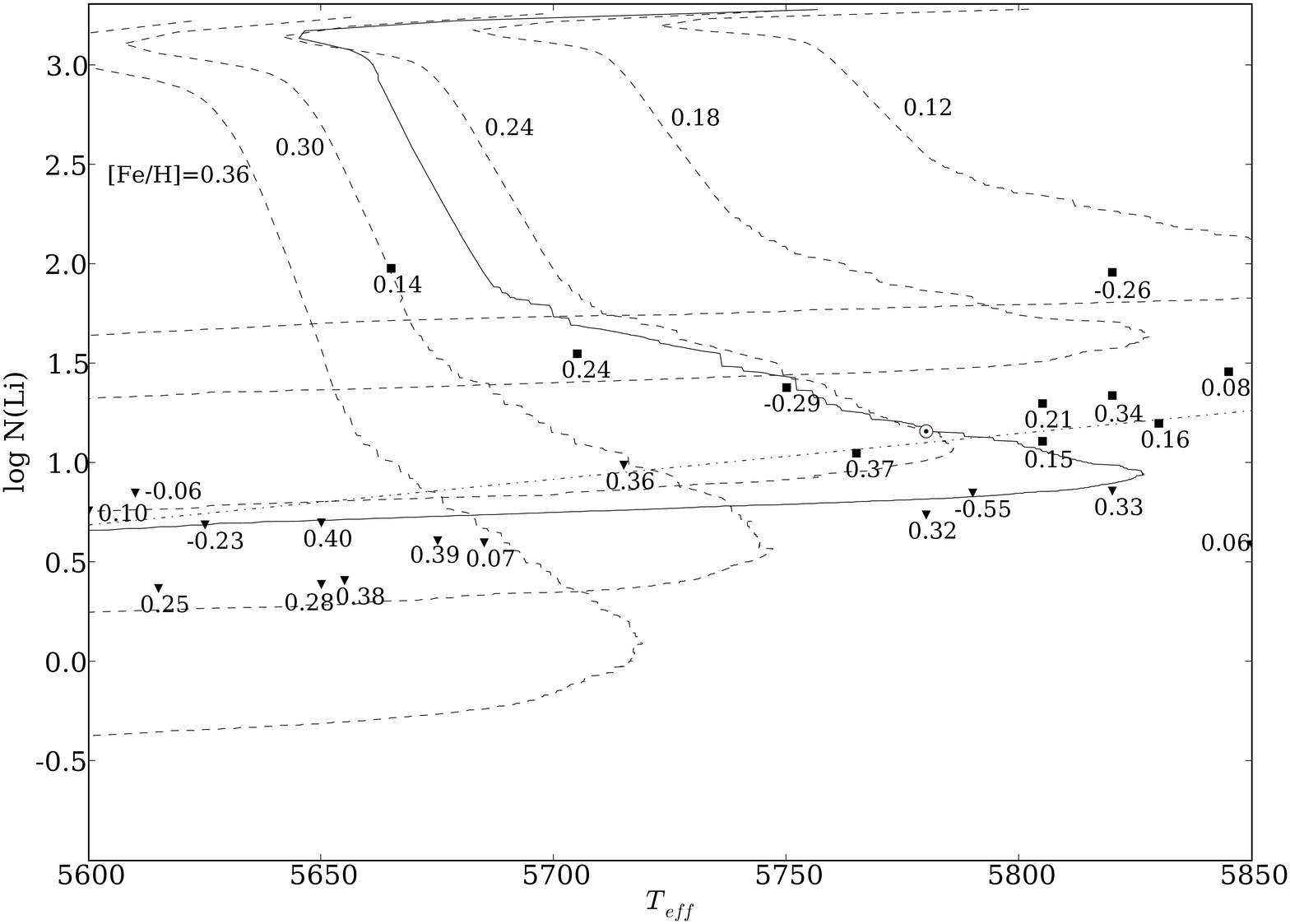}
\end{center}
\caption{Lithium destruction along the evolutionary tracks, as computed by TGEC, as a function of the effective temperature, for a standard model (solid line) of 1.00 \msol \ and overmetallic models (dashed lines) of 1.05 \msol \ and with initial metallicities [Fe/H] = 0.12, 0.18, 0.24, 0.30 and 0.36. The EHS observed by \citet{israelian04} are plotted (filled squares) and their metallicities are indicated. Upper limits are filled triangles. The position of the Sun is indicated. The dot-dashed black line is the linear regression of the lithium abundances for the observed stars.}
\label{fig:evolli-metal-sm}
\end{figure}

Figures \ref{fig:evolli-metal-sm} presents the lithium destruction along the evolutionary tracks for overmetallic models, as a function of the effective temperature. A standard model (with a solar metallicity) is used as a reference. Here we compare the effect of metallicity (for both initial overmetallicity and accretion scenarii) on models with the same mass. Overmetallic models have a mass of 1.05 \msol \ and initial metallicities [Fe/H] = 0.12, 0.18, 0.24, 0.30 and 0.36. The positions of the EHS studied in \citet{israelian04} are also plotted. For these stars (linear regression curve), the lithium depletion clearly increases for smaller effective temperatures.

 Stars with a higher surface metallicity have a deeper outer convective zone and lithium is more easily destroyed, due to the fact that the distance between the bottom of the convective zone and the nuclear destruction layer is smaller. 

The positions of the observed stars in Figure \ref{fig:evolli-metal-sm} shows that those which are cooler than the Sun can be accounted for by overmetallic models (this is also possible with other stellar masses and metallicities than showed here), whereas the stars hotter than the Sun cannot be accounted for in this framework.

\subsubsection{Influence of the stellar mass on the lithium destruction}

\begin{figure}[h!]
\begin{center}
\includegraphics[angle=0,height=8cm,width=\columnwidth]{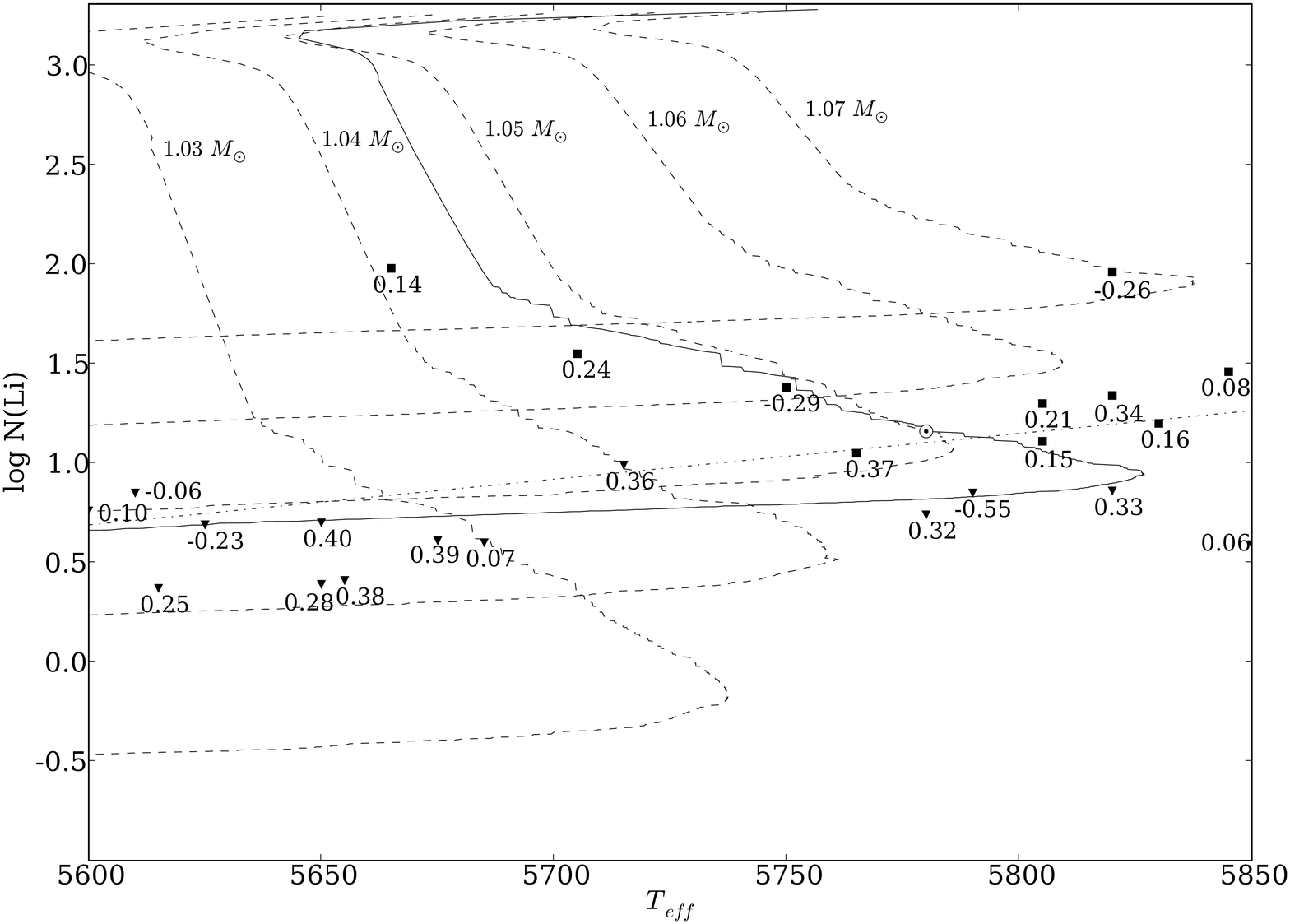}
\end{center}
\caption{Same as Figure \ref{fig:evolli-metal-sm}, but here overmetallic models with the same initial metallicity [Fe/H] = 0.24 and masses of 1.03, 1.04, 1.05, 1.06 and 1.07 \msol \ are compared.}
\label{fig:evolli-mass-sm}
\end{figure}

The lithium destruction in overmetallic models with the same surface metallicity but different masses is presented in Figure \ref{fig:evolli-mass-sm}, as a function of effective temperature. The chosen metallicity is [Fe/H] = 0.24. As previously, a standard model is given as the reference. Overmetallic models have masses of 1.03, 1.04, 1.05, 1.06 and 1.07 \msol \. \\

\begin{figure}[h!]
\begin{center}
\includegraphics[angle=0,height=8cm,width=\columnwidth]{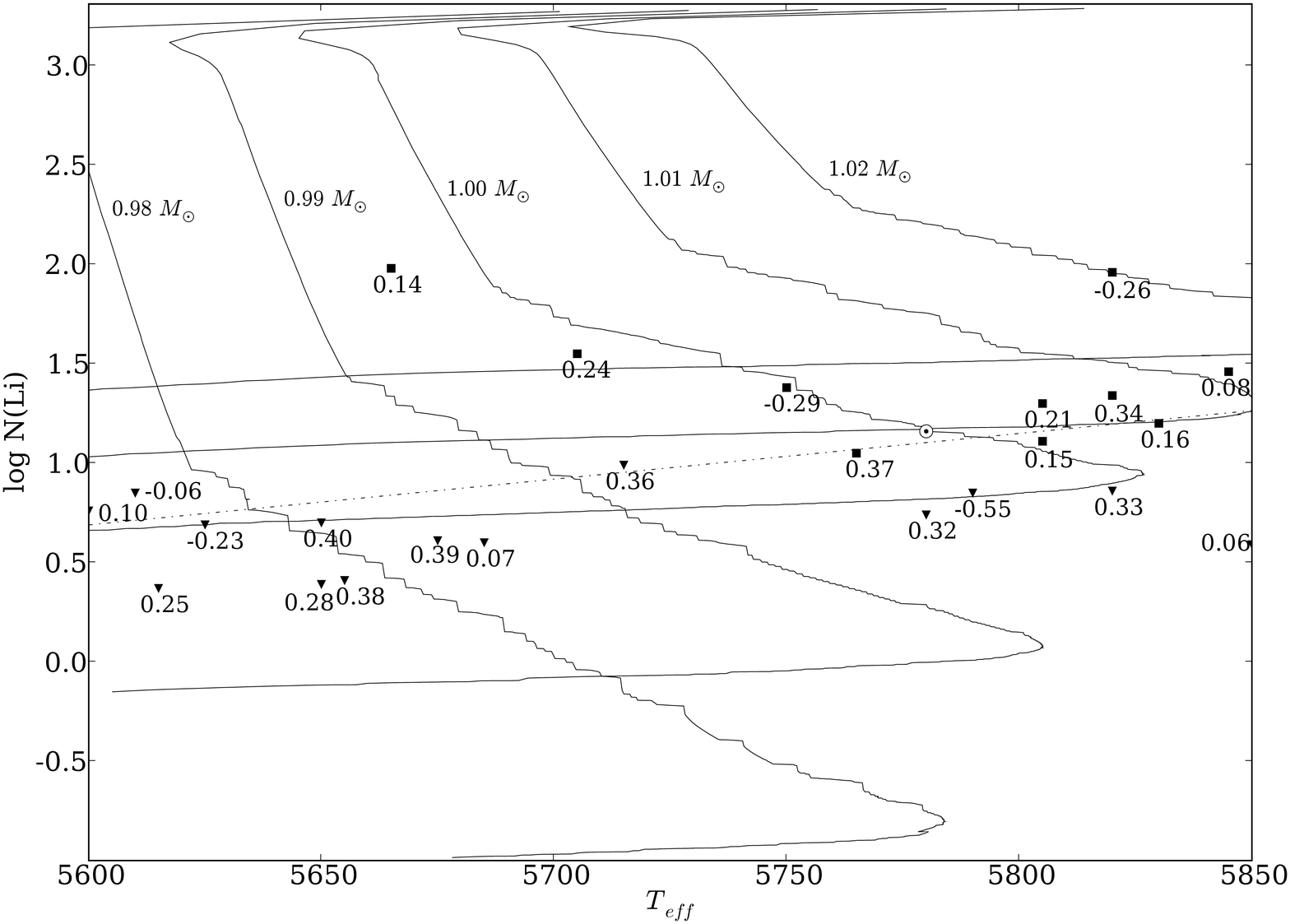}
\end{center}
\caption{Same as Figures \ref{fig:evolli-metal-sm}, but for standard models (with solar metallicity) with masses of 0.98, 0.99, 1.00, 1.01 and 1.02 \msol.}
\label{fig:evolli-stand}
\end{figure}

Figure \ref{fig:evolli-stand} presents the destruction of lithium in standard models with masses of 0.98, 0.99, 1.00, 1.01 and 1.02 \msol.  

From Figure \ref{fig:evolli-mass-sm}, we can check that the lower is the mass of the model, the more important is the lithium destruction during stellar evolution. 

As in Figure \ref{fig:evolli-metal-sm}, the hottest observed stars cannot be accounted for by overmetallic models. 

We can see from Figure \ref{fig:evolli-stand}, that the lithium destruction in these stars could be accounted for in standard models with masses between 1.00 and 1.02 \msol, but this is inconsistent with their observed overmetallicities. 

\subsection{Effective temperature scale}

The previous results depend strongly on the measures of the effective temperatures of the EHS. If the effective temperatures were lower by 50 K, which is within the error bars, the hottest observed stars would be accounted for by the models. Several works compared the effective temperature determination using spectroscopy, as the sample presented here, with the determination by the infrared flux method \citep[IRFM, see][]{blackwell80,alonso96}. \citet{ribas03}, using an IR photometry method similar to IRFM, but without requiring the use of a bolometric calibration, found a excellent agreement in the entire temperature range with the spectroscopic analysis of \citet{santos01,santos03}. The comparison done by \citet{ramirez&melendez04,ramirez&melendez05}, using IRFM, seems to imply that the temperatures derived in \citet{santos04a}, as well as by \citet{ribas03}, are hotter by about 100 K than the temperatures they obtained. In a recent work, \citet{casagrande06} show that their IRFM temperatures agree very well with spectroscopic determinations of \citet{santos04a,santos05} and \citet{luck&heiter05}. Thus, we can consider that the effective temperatures values used in this study have a good accuracy and that the possibility of systematic effects in spectroscopic metallicity determinations is small.

The uncertainties concerning the effective temperatures derived from our models are difficult to evaluate, in the same way as for all other 1-D, grey atmospheres computations. The TGEC has been tested together with six other stellar evolution codes in the Evolution and Seismic Tools Activity (ESTA) of the CoRoT mission \citep{monteiro06}. The differences for the effective temperature between the results given by all codes, using the same physics, is around one percent. In the present situation, using a more sophisticated atmosphere, or changing the internal physics could modify the derived effective temperature. However these modifications would be similar for EHS and NEHS. The present models and their parameters have been adjusted so as to precisely account for the seismic sun, and for the lithium destruction as observed in NEHS. For these reasons, we believe that the results that we obtain are significant.

\section{Extra mixing and planets migration}

Recent observations from \citet{chen&zhao06} confirm that the strong lithium destruction in EHS, observed for effective temperatures between 5600 and 5900 K does not result from the overmetallicity of these stars, but from the presence of the planets themselves. Indeed, the overmetallic NEHS do not present a strong lithium destruction, contrary to the EHS. As already evoked in \citet{israelian04}: the strong lithium destruction in these stars could be associated with a late migration of giant planets at the end of the pre-main sequence. As found from recent models \citep[e.g.][]{ida&lin08}, giant planets can still form at that time, and migrate towards the star by the so-called ``type I'' migration mechanism, which implies strong interaction with the disc. Angular momentum transfer from the planetary disc to the external outer layers of the star could induce differential rotation between the outer convective zone and the radiative zone below. This would lead to extra mixing in this region, during a short time scale, enough to destroy lithium more efficiently than in stars without planets. In the following, we show overmetallic models including such an extra mixing. \\ 

\begin{figure}[h!]
\begin{center}
\includegraphics[angle=0,height=8cm,width=\columnwidth]{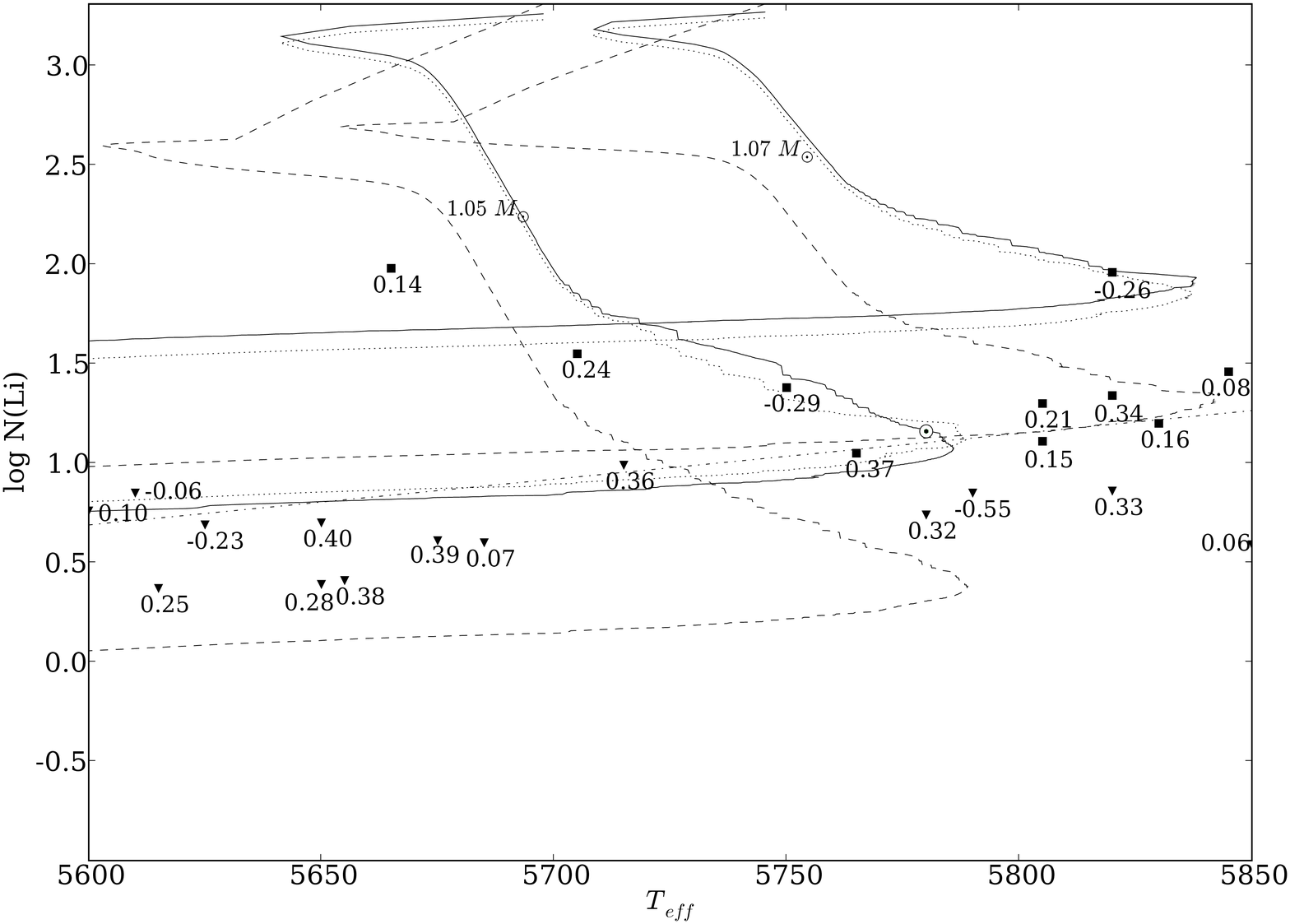}
\end{center}
\caption{Same as Figures \ref{fig:evolli-metal-sm} to \ref{fig:evolli-stand}, but for standard overmetallic models ([Fe/H]$_0$ = 0.24 and $V_i = 100$ km.s$^{-1}$) in solid line, overmetallic models ([Fe/H]$_0$ = 0.24) with initial rotation velocity $V_i = 200$ km.s$^{-1}$ in dotted line and overmetallic models ([Fe/H]$_0$ = 0.24) with an increase of the shear instability below the convective zone during the first 4 Myrs of the main sequence, in dashed line, with masses of 1.05 and 1.07 \msol.}
\label{fig:evollimigr}
\end{figure}

Figure \ref{fig:evollimigr} presents the profile of lithium destruction in overmetallic models of 1.05 and 1.07 \msol \ in different cases. The first case (continuous line) represents a star without planet, with a metallicity [Fe/H]$_0$ = 0.24 and an initial rotation velocity of 100 km.s$^{-1}$, and the mixing described in \citet{theado&vauclair03}. This model is used as a ``standard overmetallic model''.

The second case (dotted line) represents a similar model but with an initial rotation velocity of 200 km.s$^{-1}$. We can see that the increase of the initial rotation velocity has a very weak influence on the lithium depletion during the evolution. It is due to the efficient braking of the rotation, both models reaching the same rotation rate after 100 Myrs.

In the third case (dashed line) we simulate the mixing associated with a shear instability below the convective zone, due to the angular momentum transfer, by an increase of the parameters of the shear layer in the standard overmetallic model during the beginning of the main sequence. According to \citet{alibert05}, migration timescales of giant planets around the central star are found to lie between 0.1 and 10 Myrs. Their simulation of formation process of one particular planet by accretion and migration around a Sun-like star show that the object evolves from a starting distance $a_{start} = 15$ AU to a final distance $\sim 2.5$ AU from its star with a final mass of $\sim 3.5$ M$_{\rm Jup}$ in 5.5 Myrs. In our model, the diffusion coefficient at the base of the convective zone is $D_{bcz}=2.200 \times 10^6$ cm$^2$.s$^{-1}$ (instead of $D_{bcz}=2.600 \times 10^5$ cm$^2$.s$^{-1}$ in the standard overmetallic model) and the thickness of the shear layer is increased to $\Delta= 5.2 \times 10^9$ cm (instead of 6.1 $\times$ 10$^8$ cm for the standard overmetallic model), during the first 4 Myrs, corresponding to the first 4 time steps of the evolution of the model. In Figure \ref{fig:evollimigr}, the hottest observed stars can be accounted for by the model of 1.07 \msol \ with an increase of the shear instability at the beginning of the main sequence.    

The parameters that we used to simulate this extra mixing layer have been chosen to obtain results consistent with the lithium depletion of the hotter observed stars. They are somewhat arbitrary and certainly do not correspond to a unique solution. They only allow to show which kind of extra mixing is needed to explain the strong lithium depletion in cold EHS. A complete hydrodynamical treatment of the planetary-migration-induced angular momentum transfer would be needed for a stronger conclusion.

\section{Conclusion}

The purpose of this work was to present an analysis and interpretation of the observations of the lithium abundances in EHS and NEHS as given by \citet{israelian04}. We computed many evolutionary tracks with various masses and metallicities.

The first result is that our models reproduce the increase of the lithium destruction below 5900 K towards smaller effective temperature. The lithium depletion in stars cooler than the Sun, as observed by \citet{israelian04} could be accounted for in the framework of overmetallic models. However, none of these models can account for the case of hotter stars: extra mixing below the convective zone is needed to account for the observations.

We have computed specific models including such an extra mixing. They show that the observed EHS hotter than the Sun can be accounted for in overmetallic models with an increase of the shearing instability below the convective zone during the first 4 Myrs of the main sequence. However these models are very simple and contain adjusted parameters.
This result is in favour of the suggestion by \citet{israelian04} and \citet{chen&zhao06} that the strong lithium depletion observed in these stars could be related to the planet migration mechanism, which would induce an extra mixing process below the outer convective zone at the end of the pre main-sequence. A complete treatment of this process would need hydrodynamical computations of the angular momentum transfer and shear flow instabilities inside the star. However, our results show that extra mixing in an initially overmetallic star is the best scenario to account for the extra-lithium depletion observed in EHS stars. \\

Planetary migration is not the only possible explanation for the needed extra mixing in EHS. Another explanation of the extra lithium depletion in EHS stars could be related to the rotation history of the star \citep{edwards93,wolff04,wolff07}. More recently, \citet{bouvier08} has developed a consistent theory of the rotation history of solar type stars, taking into account three periods: the PMS, where the stellar rotation is magnetically coupled with the accretion disc, the approach to the ZAMS where the stellar rotation increases, and the Main Sequence relaxation, where the stars spin down, leading to similar rotation rates for all the stars at the age of the Sun. The stellar rotation rate on the ZAMS depends on the initial rotation, but also on the lifetime of the disc: for larger lifetimes, the rotation rate is smaller. To account for the observations of rotation in solar type stars from PMS to the sun, one has to assume that slowly rotating stars develop larger velocity gradients at the base of the convective envelope than rapidly rotating stars. In this framework, the extra mixing needed to account for the larger lithium depletion in EHS than in NEHS could be related to a longer lifetime of the accretion disc, possibly needed for the formation and migration of giant planets.

Such alternative explanations for the lithium observations in EHS versus non-EHS stars still present many unknowns and underlying assumptions. Although the planet migration scenario for the lithium extra depletion in EHS stars does not represent a unique possibility, the preliminary results we have obtained here are encouraging for further investigations.

\begin{acknowledgements}
Matthieu Castro and Nuno C. Santos would like to thank the support from Funda\c{c}\~ao para a Ci\^encia e a Tecnologia, Portugal, in the form of a grant (reference POCI/CTE-AST/56453/2004). This work was partially supported by the EC's FP6 and by FCT (with POCI2010 and FEDER funds), within the HELAS international collaboration.    
\end{acknowledgements}

\bibliographystyle{aa}

\end{document}